\documentclass{SciPost}

\hypersetup{
    colorlinks,
    linkcolor={red!50!black},
    citecolor={blue!50!black},
    urlcolor={blue!80!black}
}

\usepackage[bitstream-charter]{mathdesign}
\DeclareSymbolFont{usualmathcal}{OMS}{cmsy}{m}{n}
\DeclareSymbolFontAlphabet{\mathcal}{usualmathcal}

\fancypagestyle{SPstyle}{
\fancyhf{}
\lhead{\colorbox{scipostblue}{\bf \color{white} ~SciPost Physics Proceedings }}
\rhead{{\bf \color{scipostdeepblue} ~Submission }}

\fancyfoot[C]{\textbf{\thepage}}
}

\begin{document}

\pagestyle{SPstyle}

\begin{center}{\Large \textbf{\color{scipostdeepblue}{
Jevons' Paradox and Fast Generative Simulation for HEP:
Why Realistic Benchmarking is Essential \\
}}}\end{center}

Thorsten~Buss\textsuperscript{1,2},
\textbf{Henry~Day-Hall}\textsuperscript{$\star$1},
Frank~Gaede\textsuperscript{1},
Gregor~Kasieczka\textsuperscript{2},
Katja~Krüger\textsuperscript{1},
Anatolii~Korol\textsuperscript{1},
Thomas~Madlener\textsuperscript{1},
Peter~McKeown\textsuperscript{3},
Martina~Mozzanica\textsuperscript{2} and
Lorenzo~Valente\textsuperscript{2}

\begin{center}
{\bf 1} Deutsches Elektronen-Synchrotron DESY, Hamburg, Germany
\\
{\bf 2} University of Hamburg, Hamburg, Germany
\\
{\bf 3} CERN, Geneva, Switzerland
\\[\baselineskip]
$\star$ \href{mailto:email1}{\small henry.day-hall@desy.de}
\end{center}

\definecolor{palegray}{gray}{0.95}
\begin{center}
\colorbox{palegray}{
  \begin{tabular}{rr}
  \begin{minipage}{0.14\textwidth}
    \includegraphics[width=25mm]{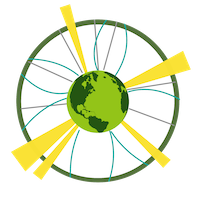}
  \end{minipage}
  &
  \begin{minipage}{0.58\textwidth}
    \vspace{5pt}
    \begin{center} \hspace{5pt}
    {\it Sustainable HEP workshop 2026 (SustHEP2026)} \\
    {\it Online, 08-10 July 2026
    }\\
    \doi{10.21468/SciPostPhysProc.?}
    \vspace{5pt}
    \end{center}
    
  \end{minipage}
\end{tabular}
}
\end{center}

\section*{\color{scipostdeepblue}{Abstract}}
\textbf{\boldmath{%
Simulation is a major computational expense in HEP, and calorimeter simulation in particular drives the overall energy cost of our physics analyses. Future detectors will contain more finely grained calorimeters than ever, and their data analyses will demand unprecedented simulated statistics.
Fast generative models redefine what is possible, producing simulations 100 times more efficiently. This article addresses Jevons' paradox in our field and considers the importance of realistic metrics in achieving ``true'' efficiency. 
}}

\vspace{\baselineskip}

\noindent\textcolor{white!90!black}{%
\fbox{\parbox{0.975\linewidth}{%
\textcolor{white!40!black}{\begin{tabular}{lr}%
  \begin{minipage}{0.6\textwidth}%
    {\small Copyright attribution to authors. \newline
    This work is a submission to SciPost Phys. Proc. \newline
    License information to appear upon publication. \newline
    Publication information to appear upon publication.}
  \end{minipage} & \begin{minipage}{0.4\textwidth}
    {\small Received Date \newline Accepted Date \newline Published Date}%
  \end{minipage}
\end{tabular}}
}}
}




\section{Introduction}\label{sec:intro}
Simulation underpins detector design, reconstruction development, and above all the comparison of predicted and observed quantities, which demands by far the largest volume of events.
LHCb records \(\mathcal{O}(10^{5})\) events per second, and keeping the statistical uncertainty from simulation subdominant requires simulating ten times as many.
Since a full Monte Carlo (MC) simulation such as Geant4~\cite{Geant4} takes \(\mathcal{O}(10)\) to \(\mathcal{O}(100)\) seconds of CPU time per photon~\cite{CaloClouds3}, and each event is likely to contain \(\mathcal{O}(1000)\) such particles, this amounts to \(\mathcal{O}(10^8)\) CPU seconds per second of data taking, even if each recorded event were relevant to only one analysis.
Sustainability considerations aside, the HEP community simply does not have that much compute.
Fast simulation techniques\footnote{In other fields such models are sometimes called surrogate models or digital twins.}, which replace the full MC with more compute-efficient methods, have therefore been in use for over a decade: first frozen showers~\cite{frozenShowers} and parametrised detector response~\cite{atlFastPCA}, then machine learning (ML), which replicates the full simulation more accurately and flexibly.
The vast volume of inference required outweighs the cost of simulating training data and running a training, so ML delivers substantial net savings; the ATLAS~\cite{ATLASfast3} and CMS~\cite{CMSfastsim} fast simulations are current production examples. 
Development remains very active, driven by new generative ML techniques and by the fidelity and luminosity expected of future detectors: following the 2025 physics briefing~\cite{physicsBriefing} which has lead to the 2026 European Strategy for Particle Physics~\cite{EUSPP}, the Tera-Z program of the planned FCC-ee is expected to deliver \(\mathcal{O}(10^{12})\) events at the \(Z\)-pole alone, with detectors \(\mathcal{O}(10^3)\) times more granular.
Generative adversarial networks, diffusion models, and Transformer attention models show great promise under these challenging conditions.

\section{Jevons' paradox and generative fast simulation}\label{sec:jevon}
Fast simulation has made individual analyses more efficient, but it has not reduced the total time spent simulating events, and so has delivered no real energy or environmental saving; the fraction of HEP computation devoted to simulation has changed little over time, while the overall budget has grown.
Some of that growth is unavoidable, as rising luminosity raises the baseline demand: the LHC's high-luminosity upgrade from \(300~\mathrm{fb}^{-1}\) to \(3000~\mathrm{fb}^{-1}\)~\cite{AtlasHighLumiPerformance} brings a corresponding \(\mathcal{O}(10)\) increase in the predicted required compute budget~\cite{AtlasComputeRoadmap}.
Luminosity, however, sets only the \emph{rate} of simulation; its quality also drives the cost, and as detectors gain fidelity the more detailed simulations offered by generative methods become increasingly attractive to use in analysis.

This second effect is Jevons' paradox: in 1865, William Stanley Jevons observed that as coal use became more efficient, total consumption \emph{increased}.
Greater efficiency improves a process's reward-to-cost ratio, making it worthwhile in more cases; if the new use cases multiply the rate of use by more than consumption per use was reduced, total consumption rises.
High quality simulation has become more efficient, yet cost effective enough that more resources than ever are spent on event simulation overall.

Fortunately, a new direction may be emerging.
It is tempting to assume that higher quality simulation always improves the accuracy of an analysis, but analysis tools have their own limits of fidelity: a highly efficient, sufficiently accurate simulation could cap the gains from simulation before the compute budget was expended.
The next sections describe such a simulation and show that its accuracy exhausts what the reconstruction can resolve.

\section{CaloClouds3}\label{sec:ccthree}
CaloClouds3 is a hybrid generative model, combining a normalising flow with a diffusion model, that simulates photons in a Higgs-factory calorimeter.
The full description and kinematic evaluation are given in the original paper~\cite{CaloClouds3}; here we summarise only the design elements that promote computational efficiency.

The first is the representation.
In the majority of photon showers only a sparse set of cells receive energy, so the shower is generated as a point cloud of energy deposits, which are aggregated into detector cells afterwards.

The second is that
two components divide responsibilities based on physical significance. Energy and active-cell counts per layer are highly correlated and informative, unlike the exact positions of individual deposits, which rarely impact identification or reconstruction. Thus, the normalizing flow predicts per-layer quantities upfront, while the diffusion model independently generates the individual points.

Because points are treated independently, inference remains efficient: a single call produces all deposit coordinates, the model vectorizes effectively, and minimal inter-point communication ensures good performance even on CPUs. Finally, the flow’s per-layer distributions are applied to the points, and their energies are assigned to cells to finalize the inference.
This is typically \(\mathcal{O}(100)\) times faster than a full MC simulation of a photon on CPU, and \(\mathcal{O}(1000)\) times faster on GPU.

\section{Constructing realistic benchmarks}\label{sec:benchmarks}
Fast simulation models are conventionally evaluated on detector-cell-level distributions known to be important to the photon shower.
Matching these is a good initial indicator of performance, but it is imperfect in two ways:
\begin{itemize}
    \item A model can match a few high-level cell-energy distributions very well and still contain artefacts that would influence reconstruction.
    \item Accuracy and computational efficiency trade off against one another; a model that perfectly replicated every cell-energy distribution would likely be slower than the Monte Carlo itself. Cell-level comparisons say nothing about what constitutes a good enough match.
\end{itemize}
Since the goal is a model that can substitute for the MC in an analysis, it must be judged in that setting: once the two agree within uncertainties \emph{after reconstruction}, they can be safely exchanged.
We therefore integrated CaloClouds3 with the analysis software chain, the Key4HEP toolkit~\cite{Key4hep}, and reproduced key reconstruction quantities as they would appear in an analysis.
These included reconstructed photon energy, energy resolution and multiplicity, alongside full \(\pi^0\) reconstruction with rates and classes of misidentification.
Here we focus on the energy and mass spectra of the reconstructed \(\pi^0\) as they are a 
good starting point for viewing a fast simulation in a realistic analysis context.
While single-photon energy deposits can be assessed directly, \(\pi^0\) decays into nearby photons often require reconstruction — to determine whether the photons are distinguishable and how their overlap affects measurements.

\section{Benchmark results}\label{sec:results}

\begin{figure}
\hspace{-0.01\textwidth}
\begin{minipage}{0.7\textwidth}
\begin{center}
    \includegraphics[width=\textwidth]{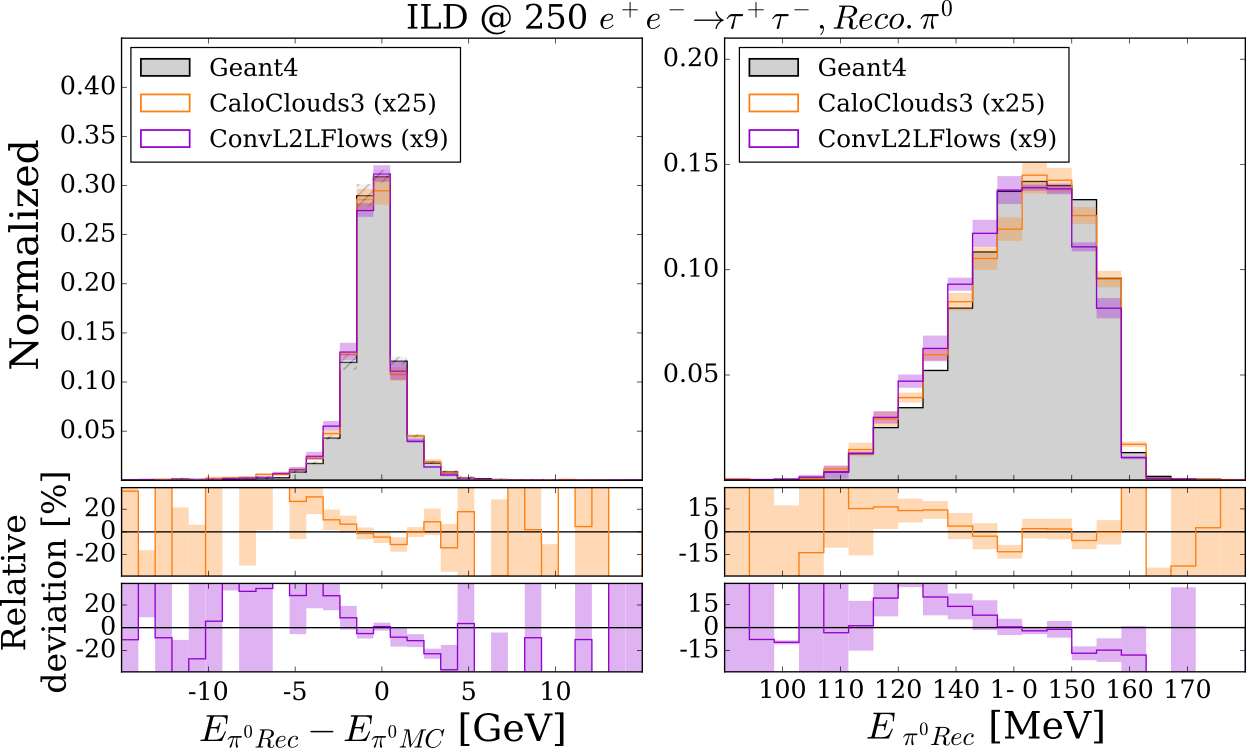}
\end{center}
\end{minipage}
\hfill
\begin{minipage}{0.3\textwidth}
\captionsetup{width=0.99\textwidth}
    \caption{
    \small{}
    The energy (left) and mass (right) of \(\pi^0\)s created in inclusive \(e^+e^- \rightarrow \tau^+\tau^-\) events at a centre-of-mass energy of 250~GeV.
        Geant4 (solid grey) forms the ground truth, against which
        two models, CaloClouds3 and ConvL2LFlows, are compared.
        Upper panels show binned counts with errors; lower panels show the ratio of each model to Geant4, with errors.
        See reference~\cite{benchmarks} for further details.
    }\label{fig:pi}
\end{minipage}
\end{figure}

The full benchmark results can be found in the dedicated publication~\cite{benchmarks}.
Here, a condensed version of a key plot is reproduced as figure~\ref{fig:pi}; it offers a good summary of all four reconstruction metrics.

The figure compares the mass and reconstructed energy of \(\pi^0\)s as simulated by a gold-standard Monte Carlo simulation (Geant4) and by two models: the highly efficient CaloClouds3 (described in section~\ref{sec:ccthree}) and, for comparison, another fast generative model, ConvL2LFlows~\cite{L2LFlows}.
CaloClouds3 matches the results of Geant4 in the majority of bins, with a number of deviations consistent with the fluctuations suggested by the error bars.
While models more accurate than CaloClouds3 can be developed, there is no practical use for that accuracy in this reconstruction.

Of course, many other reconstruction tools and analysis-level quantities exist, and exploring them with fast simulation models is an essential further validation step.

\section{Conclusion}

Jevons' paradox illustrates the surprising mechanism by which a more resource-efficient process may result in more resource consumption through its impact on user behaviour.
In the case of fast simulation, the concern is that a more computationally efficient detector simulation might be seen as a good investment for a larger share of analysis work, and therefore result in more compute used overall.
To counter this, we must appeal to other bottlenecks on the demand for analysis simulation: firstly the total observed data, and secondly the accuracy--complexity trade-off of the simulation itself.
The first is an unavoidable fixture; an analysis will aim to generate a volume of events an order of magnitude larger than the number of observed events meeting the corresponding requirements, but this number is driven by the detector luminosity and can be exhausted.
The second requires more careful handling, as the fidelity, or accuracy, of a simulation plays a significant role in its computational cost.
While it is tempting for those working on fast simulation models to match the full-scale Monte Carlo models as closely as possible, remembering that this is only desirable up to the point of compatibility after reconstruction is essential to moderating the computational cost of the whole process.
By developing appropriate benchmarks of compatibility with the baseline Monte Carlo simulation, the efficiency--accuracy trade-off can be optimised and the best possible computational efficiency of the overall simulation process guaranteed. 

\section*{Acknowledgements}
This research was supported in
part by the Maxwell computational resources operated at Deutsches Elektronen-Synchrotron DESY,
Hamburg, Germany.

\paragraph{Funding information}
This project has received funding from the European Union’s Horizon 2020
Research and Innovation programme under Grant Agreement No 101004761. We acknowledge
support by the Deutsche Forschungsgemeinschaft under Germany’s Excellence Strategy – EXC
2121 Quantum Universe – 390833306 and via the KISS consortium (05D23GU4, 13D22CH5)
funded by the German Federal Ministry of Research, Technology and Space (BMFTR) in the ErUM-Data
action plan. A.K. has received support from the Helmholtz Initiative and Networking Fund’s initiative
for refugees as a refugee of the war in Ukraine. P.M. has benefited from support by the CERN Strategic R$\&$D Programme on Technologies for Future Experiments~\cite{EPRD}.

\bibliography{SciPost_Example_BiBTeX_File.bib}

\end{document}